\documentclass[preprint,aps,showpacs,nofootinbib,tightenlines]{revtex4}
\usepackage{amsmath}
\usepackage{amssymb}
\usepackage{epsfig}
\usepackage{graphicx}
\begin{document}

\newcommand{\beq}{\begin{eqnarray}}
\newcommand{\eeq}{\end{eqnarray}}
\newcommand{\non}{\nonumber\\ }

\def\acp{{\cal A}_{CP}}
\def\calbr{ {\cal Br} }

\newcommand{\real}{{\rm Re}\,}
\newcommand{\im}{{\rm Im}\,}
\newcommand{\ov}{ \overline }
\newcommand{\calm}{ {\cal M} }

\newcommand{\psl}{ P \hspace{-2.5truemm}/ }
\newcommand{\nsl}{ n \hspace{-2.2truemm}/ }
\newcommand{\vsl}{ v \hspace{-2.2truemm}/ }
\newcommand{\epsl}{\epsilon \hspace{-1.6truemm}/\,  }
\def \etar{ \eta^\prime }
\def \etap{ \eta^{(\prime)}}

\def \epjc{ Eur. Phys. J. C }
\def \jpg{  J. Phys. G }
\def \npb{  Nucl. Phys. B }
\def \plb{  Phys. Lett. B }
\def \pr{  Phys. Rep. }
\def \rmp{ Rev. Mod. Phys. }
\def \prd{  Phys. Rev. D }
\def \prl{  Phys. Rev. Lett.  }
\def \zpc{  Z. Phys. C  }

\def \jhep{ J. High Energy Phys.  }
\def \ijmpa { Int. J. Mod. Phys. A }

\title{Revisiting $K\pi$ puzzle in the pQCD factorization approach}
\author{Wei Bai, Min Liu, Ying-Ying Fan, Wen-Fei Wang, Shan Cheng and Zhen-Jun Xiao
\footnote{Email Address: xiaozhenjun@njnu.edu.cn} }
\affiliation{Department of Physics and Institute of Theoretical Physics,\\
Nanjing Normal University, Nanjing, Jiangsu 210023, People's Republic of China}
\date{\today}
\begin{abstract}
In this paper, we calculated the branching ratios and direct CP violation
of the four $B\to K\pi$ decays with the inclusion of all currently known
next-to-leading order (NLO) contributions by employing the
perturbative QCD (pQCD) factorization approach. We found that
(a) Besides the $10\%$ enhancement from the NLO vertex corrections, the
quark-loops and magnetic penguins, the NLO contributions  to the form factors
can provide an additional $\sim 15\%$ enhancement to the branching ratios,
and lead to a very good agreement with the data;
(b) The NLO pQCD predictions are $\acp^{dir}(B^0\to K^+\pi^- )=(-6.5\pm 3.1)\%$ and
$\acp^{dir}(B^+\to K^+ \pi^0)=(2.2\pm 2.0)\%$,
become well consistent with the data due to the inclusion of the NLO contributions.
\end{abstract}


\maketitle

\section{Introduction}\label{sec:1}

The four $B \to K \pi$ decays play an important role in the precision test of
the standard model (SM) and the searching for the new physics beyond the
SM \cite{pdg2012}.
The branching ratios of these four decays have been measured with high precision
\cite{pdg2012,hfag2012}, but it is still very difficult to interpret
the so-called $``K\pi"$-puzzle:
why the measured direct CP violation $\acp^{dir}(B^0\to K^\pm \pi^\mp)$
and $\acp^{dir}(B^\pm\to K^\pm \pi^0)$ are so different ?
At the quark level, $B^0\to K^+ \pi^-$ and $B^+\to K^+\pi^0$ decay differ only
by sub-leading color-suppressed tree and the electroweak penguin.
Their CP asymmetry are expected to be similar, but the measured values differ
by $5\sigma$ \cite{pdg2012,hfag2012,exp2013}:
$ \acp^{exp}(B^0\to K^+ \pi^-) = -0.087 \pm 0.008$
while $\acp^{exp} (B^+\to K^+ \pi^0) =  0.037 \pm 0.021$.

In Ref.~\cite{nlo05}, the authors studied the $``K\pi"$ puzzle in the pQCD
factorization approach, took the NLO contributions known at 2005 into account,
and provided a pQCD interpretation for the large difference between
$\acp^{dir}(B^0\to K^\pm \pi^\mp)$ and $\acp^{dir}(B^\pm\to K^\pm \pi^0)$.
In this paper, we re-calculate these four $B \to K\pi$ decays with the inclusion
of all currently known NLO contributions in the pQCD approach, especially
the newly known NLO corrections to the form factors of $B \to (K,\pi)$
transitions \cite{prd85-074004}.

The paper is organized as follows. In Sec.II we calculate the
decay amplitudes for the considered decay modes.
The numerical results, some discussions and short summary,
are presented in Sec.III.

\section{Decay amplitudes in the pQCD approach } \label{sec:3}

In the pQCD approach, we treat the $B$ meson as a heavy-light system, and consider
the $B$ meson at rest for simplicity. By using the light-cone coordinates, the $B$
meson momentum $P_B$ and the two final state mesons' momenta $P_2$ and $P_3$
(for $M_2$ and $M_3$, respectively) can be written as
\beq
P_B = \frac{M_B}{\sqrt{2}} (1,1,{\bf 0}_{\rm T}), \quad
P_2 = \frac{M_B}{\sqrt{2}}(1-r_3^2,r^2_2,{\bf 0}_{\rm T}), \quad
P_3 = \frac{M_B}{\sqrt{2}} (r_3^2,1-r^2_2,{\bf 0}_{\rm T}),
\eeq
where $r_i^2=m_i^2/M_B^2$ are very small for $m_i=(m_\pi,m_K)$ and
will be neglected safely. Putting the light quark momenta in $B$, $M_2$
and $M_3$ meson as $k_1$, $k_2$, and $k_3$, respectively, we can choose
\beq
k_1 = (x_1 P_B^+,0,{\bf k}_{\rm 1T}), \quad
k_2 = (x_2 P_2^+,0,{\bf k}_{\rm 2T}), \quad
k_3 = (0, x_3 P_3^-,{\bf k}_{\rm 3T}).
\eeq
The decay amplitude after the integration over $k_{1,2}^-$ and $k_3^+$ can
then be written as
\beq
{\cal A}(B_d \to M_2 M_3 ) &\sim
&\int\!\! d x_1 d x_2 d x_3 b_1 d b_1 b_2 d b_2 b_3 d b_3 \non &&
\hspace{-2cm}\cdot \mathrm{Tr} \left [ C(t) \Phi_B(x_1,b_1) \Phi_{M_2}(x_2,b_2)
\Phi_{M_3}(x_3, b_3) H(x_i, b_i, t) S_t(x_i)\, e^{-S(t)} \right ],
\quad \label{eq:a2}
\eeq
where $b_i$ is the conjugate space coordinate of $k_{iT}$.
$C(t)$ is the Wilson coefficient evaluated at scale $t$,
the hard function $H(k_1,k_2,k_3,t)$ describes the four
quark operator and the spectator quark connected by  a hard gluon.
The wave function $\Phi_B(k_1)$ and $\Phi_{M_i}$ describe the hadronization of the
quark and anti-quark in the $B$ meson and $M_i$ mesons.
The Sudakov factor $S_t(x_i)$ and $e^{-S(t)} = e^{-S_B(t)-S_{M_2}(t)-S_{M_3}(t)}$
can together suppress the soft dynamics effectively \cite{li2003}.

For the B meson,  we adopt the widely used distribution amplitude
$\phi_{B}$ as in Refs.~\cite{keum01,lu01,xiao2008a}
\beq
\phi_{B}(x,b)&=& N_Bx^2(1-x)^2 \exp\left[-\frac{1}{2}\left(\frac{xm_B}{\omega_b}\right)^2
-\frac{\omega_b^2 b^2}{2}\right] \;,
\eeq
where the normalization factor $N_B$ depends on the values of the shape
parameter $\omega_B$ and the decay constant $f_B$ and defined through
the normalization relation $\int_0^1dx\; \phi_B(x,b=0)=f_B/(2\sqrt{6})$.
The shape parameter $\omega_b =0.40\pm 0.04$ has been fixed \cite{li2003}
from the fit to the $B \to \pi$ form factors derived from lattice QCD and
from Light-cone sum rule. For the light $\pi$ and $K $ mesons, we  adopt
the same set of distribution amplitudes $\phi_{\pi,K}^{A,P,T}(x_i)$  as
those defined  in Ref.~\cite{ball2006} and being used widely
for example in Refs.\cite{liy2004,ali2007,xiao2008a}.

\subsection{Leading-order contributions}

\begin{figure}[tb]
\vspace{-6cm}
\includegraphics[scale=1]{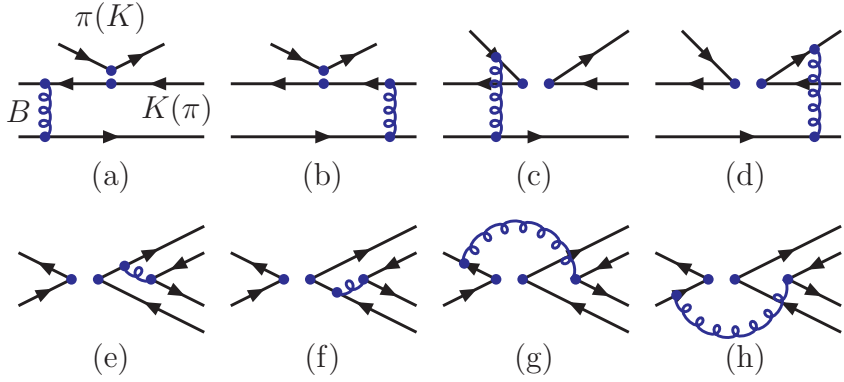}
\vspace{-19cm}
\caption{The Feynman diagrams for the LO contributions in the pQCD approach:
(a,b) factorizable emission diagrams; (c,d) hard-spectator diagrams;
(e-h) annihilation diagrams.}
\label{fig:fig1}
\end{figure}

In the pQCD factorization approach, the leading order contributions to $B \to K \pi$ decays come from
the eight Feynman diagrams as shown in Fig.1.
Following Ref.~\cite{ali2007}, we here also use the terms $(F_{e}^{LL},F_{e}^{LR},F_{e}^{SP})$ and
$(M_{e}^{LL},M_{e}^{LR},M_{e}^{SP})$ to describe the contributions
from the factorizable emission diagrams (Fig.1(a) and 1(b)) and non-factorizable emission diagrams (Fig.1(c) and 1(d))
through the $(V-A)(V-A)$, $(V-A)(V+A)$ and $(S-P)(S+P)$ operators, respectively.
In a similar way, we also adopt $(F_{a}^{LL},F_{a}^{LR},F_{a}^{SP})$
and $(M_{a}^{LL},M_{a}^{LR},M_{a}^{SP})$ to stand for the contributions
from the factorizable annihilation diagrams (Fig.1(e) and 1(f))
and non-factorizable annihilation diagrams (Fig.1(g) and 1(h)).
From the analytic calculations we obtain all relevant decay amplitudes
for the four $B \to K \pi$ decays:

By evaluating the emission diagrams Fig.1(a)-1(d), for example,
we find the following decay amplitudes
\beq
F_{e}^{LL}&=& - F_{e}^{LR}=16 \pi C_F M_B^2  \int_0^1 d x_{1} dx_3\, \int_{0}^{\infty}
b_1 db_1 b_3 db_3\; \phi_B(x_1)
\non & &
\times \left \{ \left [ (x_3+1) \phi_3^A(x_3)
+ r_3 (1-2x_3) \left ( \phi_3^P(x_3) +\phi_3^T(x_3)\right )\right ]\cdot
 h_a(x_1,x_3,b_1,b_3) \;E_e(t_a)\right.\non
&& \left.+2 r_3 \phi_3^P(x_3) \cdot
h_b(x_1,x_3,b_1,b_3) \;E_e(t_b)\right \}, \label{eq:fell}\\
F_{e}^{SP}&=& 32 \pi C_F M_B^2  \int_0^1 d x_{1} dx_3\,
\int_{0}^{\infty} b_1 db_1 b_3 db_3\; \phi_B(x_1) r_2
\non & &
\times \left \{ \left [  r_3(2+x_3) \phi_3^P(x_3)
 -r_3 x_3 \phi_3^T(x_3)+\phi_3^A(x_3)\right ]
 \cdot h_a(x_1,x_3,b_1,b_3) \cdot E_e(t_a)\right.\non
&& \left.
+2 r_3 \phi_3^P(x_3)
\cdot h_b(x_1,x_3,b_1,b_3) \cdot E_e(t_b)\right \}, \label{eq:fesp}
\eeq
\beq
M_{e}^{LL}&=& \frac{64}{\sqrt{6}} \pi C_F M_B^2 \int_0^1 dx_1 d x_{2} dx_3\,
\int_{0}^{\infty} b_1 db_1 b_2 db_2 \,\phi_{B}(x_1) \phi_2^A(x_2)
\non & &
\times \left \{ \left [ \bar{x}_2 \phi_3^A(x_3)
- x_3 r_3 \left ( \phi_3^P(x_3) -\phi_3^T(x_3)\right ) \right ] \cdot
h_c(x_i,b_1,b_2) \;E_e^{'}(t_c)\right.\non
&& \left.
+ \left [ (-x_2-x_3) \phi_3^A(x_3) + x_3 r_2 \left ( \phi_3^P(x_3)+ \phi_3^T(x_3)\right)
 \right ]\cdot
h_d(x_i,b_1,b_2) \;E_e^{'}(t_d)\right \}, \label{eq:mell}
\eeq
\beq
M_{e}^{LR}&=& \frac{64}{\sqrt{6}} \pi C_F M_B^2 \int_0^1 dx_1 d x_{2} dx_3\,
\int_{0}^{\infty} b_1 db_1 b_2 db_2 \,\phi_{B}(x_1)
\non & &
\times \left \{ \left [ \bar{x}_2 \left ( \phi_2^P(x_2) +\phi_2^T(x_2)\right ) \phi_3^A(x_3)
+x_3 r_3 \left ( \phi_3^P(x_3) +\phi_3^T(x_3)\right ) \left ( \phi_2^P(x_2) -\phi_2^T(x_2)\right )
\right.\right.\non
&& \left.\left.
+\bar{x}_2r_3 \left ( \phi_3^P(x_3) -\phi_3^T(x_3)\right ) \left ( \phi_2^P(x_2) +\phi_2^T(x_2)\right ) \right]
\cdot h_c(x_i,b_1,b_2) \;E_e^{'}(t_c)\right.\non
&& \left.
-\left [ x_2  \left ( \phi_2^P(x_2) -\phi_2^T(x_2)\right ) \phi_3^A(x_3)
+x_2 r_3 \left ( \phi_3^P(x_3) +\phi_3^T(x_3)\right )\left ( \phi_2^P(x_2) -\phi_2^T(x_2)\right )
\right.\right.\non
&& \left.\left.
+x_3 r_3 \left ( \phi_3^P(x_3) +\phi_3^T(x_3)\right ) \left ( \phi_2^P(x_2) +\phi_2^T(x_2)\right ) \right]
\cdot h_d(x_i,b_1,b_2) \;E_e^{'}(t_d)\right \}, \label{eq:melr}
\eeq
\beq
M_{e}^{SP}&=& \frac{64}{\sqrt{6}} \pi C_F M_B^2 \int_0^1 dx_1 d x_{2} dx_3\,
\int_{0}^{\infty} b_1 db_1 b_2 db_2 \,\phi_B(x_1) \phi_2^A(x_2)
\non & &
\times \left \{ \left [ (x_2-x_3-1) \phi_3^A(x_3)
+ x_3 r_3  \left ( \phi_3^P(x_3) +\phi_3^T(x_3)\right ) \right]
\cdot h_c(x_i,b_1,b_2) \;E_e^{'}(t_c)\right.\non
&& \left.
+ \left [ x_2 \phi_3^A(x_3) -x_3 r_3 \left ( \phi_3^P(x_3) -\phi_3^T(x_3)\right )
\right]\cdot h_d(x_i,b_1,b_2) \;E_e^{'}(t_d)\right \}, \label{eq:mesp}
\eeq
where $r_2=m_2/m_{B}$, $r_3=m_3/m_{B}$ and $C_F=4/3$ is a color factor.
The explicit expressions for the convolution functions
$E_{e}(t_{a,})$ and $E_a^{'}(t_{c,d})$, the hard scales
$t_{a,b,c,d}$, and the hard functions $h_{a,b,c,d}(x_i,b_i)$ can be found
in Ref.~\cite{xiao2008a}. By evaluating the annihilation diagrams Fig.1(e)-1(h)
we can find the corresponding decay amplitudes $F_{a}^{LL,LR,SP}$ and $M_{a}^{LL,LR,SP}$,
similar with those as given in Eqs.(34-38) in Ref.~\cite{fan2013}.

Taking into account the contributions from different Feynman diagrams, the total decay
amplitudes for $B^0 \to K^+ \pi^-$ and $B^+ \to K^+ \pi^0$  decays can be written explicitly as:
\beq
{\cal A}(B^0 \to K^{+} \pi^{-}) &=&   V_{ub}^* V_{us}\; \left [
f_K\; a_1 F_{e}^{LL} +  C_1 \; M_{e}^{LL} \right ]\non && - V_{tb}^*
V_{ts} \Bigl \{  f_K\left( a_4+a_{10} \right)F_{e}^{LL} + f_K( a_6+
a_8 ) \; F_{e}^{SP}  + ( C_3 +C_9 )M_{e}^{LL} \non && +( C_5 +C_7
)M_{e}^{LR} + f_B  \left[  \left(a_4 -\frac{a_{10}}{2} \right) \;
F_{a}^{LL} +\left ( a_{6} -\frac{a_8}{2} \right ) \; F_{a}^{SP}
\right] \non & & +\left ( C_{3}-\frac{C_9}{2}\right )\; M_{a}^{LL}
+\left (C_{5} -\frac{C_7}{2} \right ) M_{a}^{LR} \Bigr \}, \;
\label{eq:abspipi1}
\eeq
\beq
\sqrt{2} \; {\cal A}(B^+ \to K^{+} \pi^0) &=& V_{ub}^*
V_{us}\cdot \left \{ \left [ a_1 f_K + a_2 f_{\pi} \right]F_{e}^{LL}
+ ( C_1 + C_2) M_{e}^{LL} + a_1 f_B F_{a}^{LL} + C_1 M_{a}^{LL}
\right\} \non && \hspace{-2cm}- V_{tb}^* V_{ts}\cdot \Bigl \{ \left
( a_4+ a_{10} \right) \left (f_K F_{e}^{LL} + f_B F_{a}^{LL} \right)
+(a_6 + a_8 ) \left ( f_K F_{e}^{SP} +f_B F_{a}^{SP} \right) \non &&
+ ( C_3 +C_9 ) \left ( M_{e}^{LL} +M_{a}^{LL} \right ) +( C_5 + C_7
) \left ( M_{e}^{LR} +M_{a}^{LR} \right ) \non && +\frac{3}{2}( -a_7
+ a_{9} )  f_{\pi}\; F_{e}^{LL} +\frac{3}{2} C_8 \; M_{e}^{SP} +
\frac{3}{2} C_{10} \; M_{e}^{LL} \Bigr \}, \;
\label{eq:abspipi2}
\eeq
where $a_i$ is the combination of the Wilson coefficients $C_i$ with the definitions:
$a_{1,2}=C_{2,1}+\frac{C_{1,2}}{3}$, $a_i=C_i+\frac{C_{i+1}}{3}$ 
( $a_i= C_i+\frac{C_{i-1}}{3}$)
for $i=3,5,7,9$ ( $i=4,6,8,10$) respectively.
The explicit expressions for $B^0 \to K^0 \pi^0$ and $B^+ \to K^0 \pi^+$
decays are similar with those as shown in Eqs.(\ref{eq:abspipi1},\ref{eq:abspipi2}).

\subsection{NLO contributions}


\begin{figure}[tb]
\vspace{-6cm}
\includegraphics[scale=1]{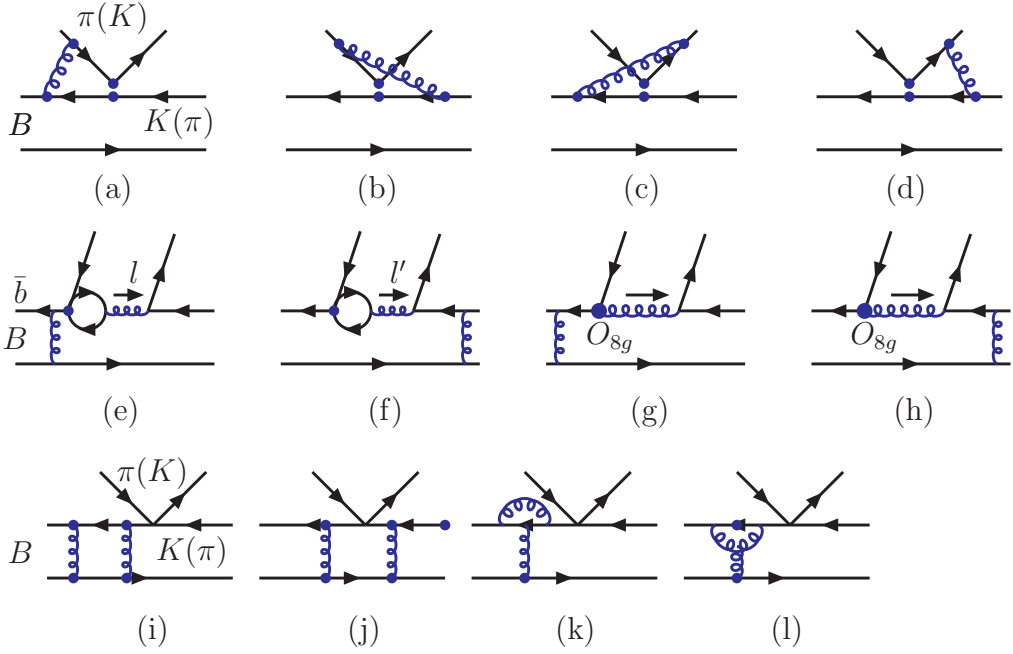}
\vspace{-15cm}
\caption{The typical Feynman diagrams for currently known NLO contributions:
the vertex corrections (a-d); the quark-loop (e-f);  the chromo-magnetic
penguins (g-h); and the NLO contributions to form factors (i-l).}
\label{fig:fig2}
\end{figure}

Based on the power counting rule in the pQCD factorization approach \cite{nlo05},
the following NLO contributions should be included\cite{nlo05}:
\begin{enumerate}
\item[(1)]
The Wilson coefficients $C_i(M_W)$ at NLO level \cite{buras96},
the renormalization group evolution matrix $U(t,m,\alpha)$ at NLO level
and the strong coupling constant $\alpha_s(t)$ at two-loop level\cite{pdg2012}.

\item[(2)]
The currently known NLO contributions to hard kernel $H^{(1)}(\alpha_s^2)$ include
\cite{nlo05,o8g2003,prd85-074004}:
\begin{itemize}
\item[(a)] The vertex correction (VC)from the Feynman diagrams Fig.2(a)-2(d);
\item[(b)] The NLO contributions from the quark-loops (QL) as shown in Fig.2(e)-2(f);
\item[(c)] The NLO contributions from the operator $O_{8g}$ as shown in Fig.3(g)-3(h)
\cite{o8g2003};
\item[(d)] The NLO contributions to the form factors as shown in
Fig.2(i)-2(l)~\cite{prd85-074004}.
\end{itemize}
\end{enumerate}

The still missing NLO parts in the pQCD approach are the $O(\alpha_s^2)$
contributions from hard spectator diagrams and annihilation diagrams,
as illustrated by Fig.5 in Ref.~\cite{fan2013}.
According to the general arguments as presented in Ref.~\cite{nlo05} and
explicit numerical comparisons of the contributions from
different sources for $B \to K \pi$ decays as made in Ref.~\cite{fan2013}
one generally believe that
these still missing NLO parts should be very small and can be neglected safely. The
major reasons are the following:
\begin{enumerate}
\item
For the non-factorizable spectator diagrams in Fig.1(c)-1(d),
their LO contributions are strongly suppressed by the isospin symmetry
and color-suppression with respect to the factorizable emission diagrams Fig.1(a)-1(b).
The NLO contributions from Figs.5(a)-5(d) in Ref.~\cite{fan2013} are higher
order corrections to small LO quantities.

\item
For the annihilation spectator diagrams at leading order, i.e. Figs.1(e)-1(h),
they are power suppressed and generally much smaller with respect to the
contributions from the emission diagrams Fig.1(a)-1(b).
The NLO contributions from Figs.5(e)-5(h) in Ref.~\cite{fan2013}
are also the higher order corrections to the small LO quantities.

\item
Taking $B^+ \to K^+ \eta$ decay as an example, as shown in Eq.(87) of
Ref.~\cite{fan2013}, the relative strength of the individual LO contribution
$\calm^{a+b}$ from the emission diagrams, $\calm^{c+d}$ and $\calm^{anni}$ from
the spectator and the annihilation diagram respectively can be evaluated through
the following ratio:
\beq
|\calm^{a+b}|^2:|\calm^{c+d}|^2:|\calm^{anni}|^2=3.23:0.02:0.33.
\eeq
One can see directly from the above ratio that the contribution from
emission diagram is indeed dominant, while the contribution from
$\calm^{c+d}$ ( $\calm^{anni}$ ) is less than $1\%$ ($10\%$) of the dominant one.

\end{enumerate}
Based on about reasonable arguments and explicit numerical examinations, one can
see that the still missing NLO parts in the pQCD approach are higher order corrections
to those small LO quantities, and therefore should be very small and can be neglected
safely. For more details of numerical comparisons, one can see
Ref.~\cite{fan2013}.


The vertex corrections from the Feynman diagrams as shown in Figs.~2(a)-2(d),
have been calculated years ago in the QCD factorization
appeoach\cite{bbns99,npb675}.
Since there is no end-point singularity in the evaluations of
Figs.2(a)-2(d), it is unnecessary to employ the $k_{T}$ factorization
theorem here \cite{nlo05}.
The NLO vertex corrections will be included by adding a same vertex function $V_i(M)$
to the corresponding Wilson coefficients $a_i(\mu)$ as in Refs.~\cite{bbns99,npb675,xiao2008a}.

For the $b\rightarrow s$ transition, the contributions from the various quark
loops are given by\cite{nlo05}
\beq
H_{eff}=-\sum_{q=u,c,t}\sum_{q^{'}}\frac{G_{F}}{\sqrt{2}}V_{qb}V_{qs}^{*}
\frac{\alpha_{s}(\mu)}{2\pi}C^{(q)}(\mu,l^{2})
\left [ \bar{s}\gamma_{\rho} \left ( 1-\gamma_{5} \right )T^{a}b \right ]
\left (\bar{q}^{'}\gamma^{\rho}T^{a}q^{'} \right ),
\eeq
where $l^{2}$ is the invariant mass of the gluon, which attaches the quark loops
in Figs.2e and 2f. The expressions of the functions $C^{q}(\mu,l^{2})$ for $q=(u,c,t)$
can be found easily in Refs.\cite{nlo05,xiao2008a}.

The magnetic penguin is another kind penguin correction induced by the insertion of
the operator $O_{8g}$, as illustrated by Fig.2(g) and 2(h). The corresponding weak
effective Hamiltonian contains the $b\to s g$ transition can be written as
\beq
H_{eff}^{mp} =-\frac{G_F}{\sqrt{2}} \frac{g_s}{8\pi^2} m_b\; V_{tb}V_{ts}^*\; C_{8g}^{eff}
\left [ \bar{s}_i \; \sigma^{\mu\nu}\; (1+\gamma_5)\;
 T^a_{ij}\; G^a_{\mu\nu}\;  b_j \right ],
\label{eq:o8g}
\eeq
where $i,j$ are the color indices of quarks,  $C_{8g}^{eff}= C_{8g} + C_5$ \cite{nlo05}
is the effective Wilson coefficient.

For the sake of convenience we denote all current known NLO  contributions except for
those to the form factors by the term Set-A.
For the four $B\to K \pi$ decays, the Set-A NLO contributions will be included in
a simple way:
\beq
{\cal A}_{\pi K} &\to & {\cal A}_{\pi K} +
\sum_{q=u,c,t} \xi_q{\cal M}_{\pi K}^{(q)} + \xi_t{\cal M}_{\pi K}^{(g)}, \non
{\cal A}_{K \pi} &\to & {\cal A}_{K \pi}
+ \sum_{q=u,c,t} \xi^{\prime}_q{\cal M}_{K \pi}^{(q)}+ \xi^{\prime}_t{\cal M}_{K \pi}^{(g)},
\eeq
where $\xi_q = V_{qb}V_{qd}^*$, $\xi^{\prime}_q = V_{qb}V_{qs}^*$ with
$q=u,c,t$, while the decay amplitudes ${\cal M}^{(q)}_{M_i,M_j}$ and
${\cal M}^{(g)}_{M_i,M_j}$ are of the form:
\beq
{\cal M}^{(q)}_{\pi^- K^+}&=&
-8m_{B}^4\frac{{C_F}^2}{\sqrt{2N_c}} \int_0^1 dx_1dx_2dx_3
\int_0^\infty b_1db_1b_3db_3 \,\phi_{B}(x_1)
\left \{ \left [ (1+x_3)\phi_{\pi}^A(x_3) \phi_K^A(x_2)
\right.\right.
\non && \left.\left.
+ 2r_{\pi}\phi_{K}^P(x_2) \phi_{\pi}^A(x_3)+
r_{\pi}(1-2x_3)\phi_{K}^{A}(x_2)(\phi_{\pi}^P(x_3)+\phi_{\pi}^T(x_3))
\right.\right.
\non && \left.\left.
+ 2r_{\pi} r_K \phi_{K}^P(x_2) ((2+x_3)\phi_{\pi}^P(x_3)
-x_3\phi_{\pi}^T(x_3)) \right ]
\right.
\non && \left.
\cdot \alpha_s^2(t_a) h_e(x_1,x_3,b_1,b_3)
\exp[-S_{ab}(t_a)] C^{(q)}(t_a,l^2)
\right.
\non && \left.
+ \left [ 2r_{\pi}\phi_{K}^A(x_2)\phi_{\pi}^P(x_3) +
4r_{\pi}r_K\phi_{K}^P(x_2)\phi_{\pi}^P(x_3) \right ]
\right.
\non && \left.
\cdot \alpha_s^2(t_b) h_e(x_3,x_1,b_3,b_1)
 \exp[-S_{ab}(t_b)] C^{(q)}(t_b,l'^2)\right \},
\label{eq:mqkpi}
\eeq
\beq
{\cal M}^{(g)}_{\pi^- K^+} &=&-16m_{B}^6\frac{{C_F}^2}{\sqrt{2N_c}} \int_0^1 dx_1dx_2dx_3
\int_0^\infty b_1db_1b_2db_2b_3db_3\, \phi_{B}(x_1)
\non &&
\cdot \left \{ \left [ (1-x_3) \left [ 2\phi_{\pi}^A(x_3) +
r_{\pi}(3\phi_{\pi}^P(x_3) +\phi_{\pi}^T(x_3) )
\right.\right.\right.
\non && \left.\left.\left.
+ r_{\pi} x_3(\phi_{\pi}^P(x_3)-\phi_{\pi}^T(x_3)) \right ]
\phi_{K}^A(x_2)
\right.\right.
\non && \left.\left.
- r_{K}x_2(1+x_3) (3\phi_{K}^P(x_2) -\phi_{K}^T(x_2))\phi_{\pi}^A(x_3)
\right.\right.
\non && \left.\left.
-r_{\pi}r_K(1-x_3)(3\phi_{K}^P(x_2)
+ \phi_{K}^T(x_2))(\phi_{\pi}^P(x_3) -\phi_{\pi}^T(x_3))
\right.\right.
\non && \left.\left.
- r_{\pi}r_K x_2 (1-2x_3)(3\phi_{K}^P(x_2) -
\phi_{K}^T(x_2))(\phi_{\pi}^P(x_3) + \phi_{\pi}^T(x_3)) \right ]
\right.
\non &&\left.
\cdot \alpha_s^2(t_a)
h_g(x_i,b_i) \exp[-S_{cd}(t_a)] C_{8g}^{eff}(t_a)
\right.
\non &&\left.
+ \left [ 4r_{\pi}\phi_{K}^A(x_2)\phi_{\pi}^P(x_3)
+ 2r_{K}r_{\pi} x_2(3\phi_{K}^P(x_2)
- \phi_{K}^T(x_2))\phi_{\pi}^P(x_3)\right ]
\right.
\non &&\left.
\alpha_s^2(t_b) h'_g(x_i,b_i)
\exp[-S_{cd}(t_b)] C_{8g}^{eff}(t_b)\right\},
\label{eq:mgkpi}
\eeq
\beq
\sqrt{2}{\cal M}^{(q)}_{K^0 \pi^0}&=&  {\cal M}^{(q)}_{\pi^- K^+}= {\cal M}^{(q)}_{K^0 \pi^+}
= {\cal M}^{(q)}_{K^+ \pi^0}, \\
\sqrt{2}{\cal M}^{(g)}_{K^0 \pi^0}&=& {\cal M}^{(g)}_{\pi^- K^+}={\cal M}^{(g)}_{K^0 \pi^+}
= {\cal M}^{(g)}_{K^+ \pi^0} ,
\eeq
where the expressions of the Sudakov factors $S_{ab}(t_i)$ and $S_{cd}(t_i)$,
the functions $C^{(q)}(t_a,l^2)$ and $C^{(q)}(t_b,l'^2)$, can be found easily in
Refs.~\cite{nlo05,xiao2008a}.

In Ref.~\cite{prd85-074004}, the authors derived the $k_{\rm T}$-dependent NLO hard kernel
$H^{(1)}$ for the $B \to \pi$ transition form factor.
Here we quote their results directly, and extend the expressions to the $B \to K$
transitions under the assumption of $SU(3)$ flavor symmetry.
At the NLO level, the hard kernel function $H$ can then be written as
\beq
H=H^{(0)}(\alpha_s)+ H^{(1)}(\alpha_s^2)=
\left [ 1+F(x_1,x_3,\mu,\mu_f,\eta,\zeta_1) \right ] H^{(0)}(\alpha_s),
\eeq
where the expression of the NLO factor $F(x_1,x_3,\mu,\mu_f,\eta,\zeta_1)$ can
be found in Eq.~(56) of Ref.~\cite{prd85-074004}.


\section{Numerical Results and Discussions}\label{sec:4}

In numerical calculations, the following input parameters will be used\cite{pdg2012}
( all the masses, QCD scale and decay constants are in units of $GeV$):
\beq
\Lambda_{QCD}&=& 0.25\; , \quad m_W = 80.40\;, \quad m_{B} = 5.28\;, \quad
m_{\pi}=0.14, \quad m_{K}=0.494 \; ;  \non
f_{\pi}&=& 0.13, \quad f_K=0.16, \quad \tau_{B^0}= 1.528\;ps,\quad \tau_{B^{+}} = 1.643\;ps.
\label{eq:mass}
\eeq
For the CKM  matrix elements in the Wolfenstein
parametrization, we use $\lambda = 0.2254$, $A= 0.817$,
$\bar{\rho} = 0.136^{+0.019}_{-0.018}$ and $ \bar{\eta}= 0.348\pm 0.013$ \cite{pdg2012}.
For the Gegenbauer moments and other relevant input parameters,
we use \cite{ball2006}
\beq
a_1^\pi &=& 0, \quad a_1^K = 0.06, \quad a_2^{\pi}=a_2^K=0.25\pm 0.15,
\quad  a_4^{\pi} = -0.015, \quad a_4^K=0,\non
\rho_\pi &=& m_\pi/m_0^\pi, \quad \rho_K = m_K/m_0^K,\quad
\eta_3=0.015, \quad \omega_3=-3.0,
\label{eq:pa-1}
\eeq
with the chiral mass $m_0^\pi=1.4 \pm 0.1$ GeV, and $m_0^K=1.6 \pm 0.1$ GeV.

From the decay amplitudes and the input parameters,
it is straightforward to calculate the branching
ratios and CP violating asymmetries for the four considered $B \to K \pi$
decays \cite{nlo05,xiao2008a}.

In Table I and II, we show the LO and NLO pQCD predictions for the branching ratios
and the direct CP violating asymmetries of the considered four $B\to K\pi$ decays.
In Table I and II, we list only the central values of the LO pQCD predictions in
column two, and the central values and the major theoretical errors simultaneously
in column four. The first error arises from the uncertainty of
$\omega_B=0.40\pm 0.04$ GeV, the second one from the uncertainty of
$a_2^{\pi,K}=0.25\pm 0.15$, and the third one is induced by the
variations of both $m_0^K=1.6 \pm 0.1$ GeV and $m_0^\pi=1.4 \pm 0.1$ GeV.
The errors induced by the uncertainties of other input parameters are very
small and have been neglected. As a comparison, we also show the partial
pQCD predictions obtained in this work ( labeled by Set-A in column three )
and those as given in Ref.~\cite{nlo05} in the column five, where the same
Set-A NLO contributions are included.  One can see from those numerical results that:
\begin{enumerate}
\item
For branching ratios, the central values of pQCD predictions as given in column three
in Table I are smaller than those as shown in column five by about thirty percent,
such difference are largely induced by the change of the lower cutoff of the hard scale
$t$ from $\mu_0=0.5$ GeV in Ref.~\cite{nlo05} to $\mu_0=1$ GeV here,
because it may be conceptually incorrect to evaluate the Wilson coefficients
at scales down to 0.5 GeV \cite{xiao2008a,beneke07}.
For direct CP violating asymmetries, as shown in the third and fifth column of
Table II,  the changes of the pQCD predictions due to the variation of $\mu_0$ are
rather small, this is consistent with the general expectation.

\item
Analogous to the case for $B \to K \etap$ decays
as shown explicitly in Table VIII and IX in Ref.~\cite{fan2013},
the NLO contributions to the decay amplitudes from the vertex, the
quark-loop and the magnetic penguins are largely canceled from each
other, and in turn leaving only a roughly $10\%$ enhancement to the
LO pQCD predictions of the branching ratios.

\item
As listed in Table I of Ref.~\cite{wang2012}, the NLO contribution to the
form factor for $B \to \pi$ ($B\to K$) transition can provide a $18\%$ ($15\%$)
enhancement to the corresponding LO result:
\beq
F_0^{LO}(0)(B\to \pi)&=&0.22\pm 0.04 \longrightarrow F_0^{NLO}(0)(B\to \pi)=0.26\pm 0.04, \non
F_0^{LO}(0)(B\to K)&=& 0.27\pm 0.05 \longrightarrow F_0^{NLO}(0)(B\to K)=0.31\pm 0.05.
\eeq
Such enhancement to form factors $F_0^{B\to \pi}(0)$ and $F_0^{B\to K}(0)$
can in turn result in an additional $12\%$ to $18\%$ enhancement to branching ratios
relative to the results in the third column with the label "Set-A",
as illustrated clearly by the numerical results in column four of Table I,
and consequently lead to a very good agreement
between the NLO pQCD predictions and the measured values within errors.

\item
For $\acp^{dir}(B^0\to K^{0}\pi^{0})$ and $\acp^{dir}(B^+\to K^{0}\pi^{+})$,
the  pQCD predictions agree well with the data.

\item
At the leading order, the pQCD predictions for
$\acp^{dir}(B^0\to K^+\pi^- )$ and $\acp^{dir}(B^+\to K^+ \pi^0)$
are indeed similar in both the sign and the magnitude, $-12.6\%$ vs $-8.6\%$,
as generally expected.
After the inclusion of the NLO contributions, however, they become rather different
as can be seen from Table II. The NLO pQCD predictions, consequently,
become agree well with the data.
One can also see that the pQCD predictions for $\acp^{dir}(B^0\to K^+\pi^- )$ and
$\acp^{dir}(B^+\to K^+ \pi^0)$ remain basically unchanged when the NLO
corrections to the form factors are taken into account.

\end{enumerate}

\begin{table}[tb]
\begin{center}
\caption{ The LO and NLO pQCD predictions for branching ratios $Br(B \to K\pi)$
(in units of $10^{-6}$ ), the previous pQCD predictions in Ref.~\cite{nlo05} and
the relevant data \cite{pdg2012,hfag2012} also be listed in last two columns.}
\label{tab:br}
\begin{tabular}{l|c|c|c|c|l}  \hline \hline
Decay modes & LO  & Set-A & NLO: This Work   & pQCD\cite{nlo05} &Data \\ \hline
  $B^0\rightarrow K^{0}\pi^{0}$               & $6.3$   & $6.6$   &$7.4^{+2.2+1.3+0.9}_{-1.5-1.2-0.9}$
  &$ 9.1^{+5.6}_{-3.3}$&$9.9\pm 0.5$ \\
\hline
  $B^0\rightarrow K^{+}\pi^{-}$                & $14.4$   & $15.3$   &$17.7^{+5.5+2.6+2.0}_{-3.8-2.4-2.0}$
  &$20.9^{+15.6}_{-6.3}$& $19.6\pm0.5$ \\
\hline
  $B^+\rightarrow K^{+}\pi^{0}$                & $10.1$   & $10.6$   &$12.5^{+4.0+1.7+1.3}_{-2.8-1.6-1.2}$
  &$13.9^{+10}_{-5.6}$&$12.9\pm0.5$ \\
\hline
  $B^+\rightarrow K^{0}\pi^{+}$             & $17.5$   & $18.4$  & $21.5^{+6.7+3.4+2.8}_{-4.7-3.1-2.3}$
  &$24.5^{+13.6}_{-8.1}$&$23.8\pm 0.7$\\
\hline\hline
\end{tabular} \end{center}
\end{table}

\begin{table}[tb]
\begin{center}
\caption{ The same as in Table I, but for the pQCD predictions for the
direct CP violations $\acp^{dir}(B \to K \pi)$ (in units of $10^{-2}$). }
\label{tab:acpdir}
\begin{tabular}{l|c|c|c|c|l}  \hline\hline
Decay modes  & LO   & Set-A  & NLO: This Work &pQCD\cite{nlo05} &Data \\\hline
\hline
$\acp^{dir}(B^0\to K^{0}\pi^{0})$           & $-2.2$     &$-7.0$
  &$-7.9^{+0.3+0.8+0.4}_{-0.23-0.9-0.5}$     &$-7\pm 3$& $0\pm 13$ \\
  $\acp^{dir}(B^+\to K^{0}\pi^{+})$           & $-0.75$     & $0.40$
  &$0.38 ^{+0.09+0.02+0.03} _{-0.11-0.07-0.05}$             &$0\pm 0$&$-1.5\pm 1.2$\\ \hline
  $\acp^{dir}(B^0\to K^{+}\pi^{-})$           & $-12.6$    & $-6.4$
  &$-6.5 ^{+2.1} _{-2.0}\pm 2.3\pm 0.3$             &$-9^{+6}_{-8}$&$-8.7\pm 0.8$ \\
  $\acp^{dir}(B^+\to K^{+}\pi^{0})$           & $-8.6$     & $2.0$
  &$2.2^{+1.7}_{-1.8} \pm 1.2 \pm 0.1$      &$-1^{+3}_{-5}$&$ 3.7\pm 2.1$ \\
\hline\hline
\end{tabular} \end{center}
\end{table}

In summary, we studied the $B \to K \pi$ decays by employing the pQCD
factorization approach. We focus on checking the effects of all currently known
NLO contributions to the branching ratios and direct CP violations of the considered
decay modes, especially the rule of the NLO corrections to
the form factors $F_0^{B\to \pi}(q^2)$ and $F_0^{B\to K}(q^2)$.
Based on the numerical calculations and the
phenomenological analysis, the following points have been observed:
\begin{enumerate}
\item
Besides the  $10\%$ enhancement from the Set-A NLO contributions,
the NLO contributions to the form factors can provide an additional $\sim 15\%$
enhancement to the branching ratios, and lead to a very good agreement with the data.

\item
With the inclusion of all known NLO contributions, the NLO pQCD
predictions are
\beq
\acp^{dir}(B^0\to K^+\pi^- )=(-6.5\pm 3.1)\%, \quad
\acp^{dir}(B^+\to K^+ \pi^0)=(2.2\pm 2.0)\%,
\eeq
where the theoretical errors have been added in quadrature, which
agree well with the data.

\end{enumerate}

\begin{acknowledgments}
This work is supported by the National Natural Science
Foundation of China under the Grant No.~10975074 and~11235005.

\end{acknowledgments}


\end{document}